\documentclass[aps,jcp,amsmath,amssymb,reprint]{revtex4-2}

\usepackage{chemformula} % Formula subscripts using \ch{}
\usepackage[T1]{fontenc} % Use modern font encodings
\usepackage{multirow}
\usepackage{dcolumn}
\usepackage{tabularx}
\usepackage{float}
\usepackage{xr}
\usepackage{hyperref}
\usepackage[utf8]{inputenc}
\usepackage{mciteplus}
\usepackage{subfig}
\usepackage[utf8]{inputenc}
\setcitestyle{super}
\usepackage[cdot,mediumqspace,amssymb]{SIunits}
\usepackage{amsmath}

\usepackage{textcomp} % to get a tilde
\newcommand{\comment}[1]{}

\begin{document}
	\preprint{AIP/123-QED}
	
	\title{High-dimensional frequency conversion in hot atomic system}
	
	\author{Wei-Hang Zhang}
	\affiliation{CAS Key Laboratory of Quantum Information, University of Science and Technology of China, Hefei 230026, China}
	\affiliation{CAS Center for Excellence in Quantum Information and Quantum Physics, University of Science and Technology of China, Hefei 230026, China}

	\author{Ying-Hao Ye}
	\affiliation{CAS Key Laboratory of Quantum Information, University of Science and Technology of China, Hefei 230026, China}
	\affiliation{CAS Center for Excellence in Quantum Information and Quantum Physics, University of Science and Technology of China, Hefei 230026, China}

	\author{Lei Zeng}
	\affiliation{CAS Key Laboratory of Quantum Information, University of Science and Technology of China, Hefei 230026, China}
	\affiliation{CAS Center for Excellence in Quantum Information and Quantum Physics, University of Science and Technology of China, Hefei 230026, China}
	
	\author{En-Ze Li}
	\affiliation{CAS Key Laboratory of Quantum Information, University of Science and Technology of China, Hefei 230026, China}
	\affiliation{CAS Center for Excellence in Quantum Information and Quantum Physics, University of Science and Technology of China, Hefei 230026, China}

	\author{Jing-Yuan Peng}
	\affiliation{CAS Key Laboratory of Quantum Information, University of Science and Technology of China, Hefei 230026, China}
	\affiliation{CAS Center for Excellence in Quantum Information and Quantum Physics, University of Science and Technology of China, Hefei 230026, China}
	
	\author{Dong-Sheng Ding}
	\email{dds@ustc.edu.cn}
	\affiliation{CAS Key Laboratory of Quantum Information, University of Science and Technology of China, Hefei 230026, China}
	\affiliation{CAS Center for Excellence in Quantum Information and Quantum Physics, University of Science and Technology of China, Hefei 230026, China}
	
	\author{Bao-Sen Shi}
	\email{drshi@ustc.edu.cn}
	\affiliation{CAS Key Laboratory of Quantum Information, University of Science and Technology of China, Hefei 230026, China}
	\affiliation{CAS Center for Excellence in Quantum Information and Quantum Physics, University of Science and Technology of China, Hefei 230026, China}

%% The abstract environment will automatically gobble the contents
%% if an abstract is not used by the target journal.
%%%%%%%%%%%%%%%%%%%%%%%%%%%%%%%%%%%%%%%%%%%%%%%%%%%%%%%%%%%%%%%%%%%%%
\begin{abstract}

One of the major difficulties in realizing a high-dimensional frequency converter for conventional optical vortex (COV) stems from the difference in ring diameter of COV modes with different topological charge numbers $l$. Here, we implement a high-dimensional frequency convertor for perfect optical vortex (POV) modes with invariant size through the four-wave mixing (FWM) process by utilizing Bessel-Gaussian beams instead of Laguerre-Gaussian beams. The measured conversion efficiency from 1530 nm to 795 nm is independent of $l$ at least in subspace $l\in\{-6,...,6\}$, and the achieved conversion fidelities for two-dimensional (2D) superposed POV states exceed 97$\%$. We further realize the frequency conversion of 3D, 5D and 7D superposition states with fidelities as high as 96.70$\%$, 89.16$\%$ and 88.68$\%$, respectively. The reported scheme is implemented in hot atomic vapor, it's also compatible with the cold atomic system and may find applications in high-capacity and long-distance quantum communication.

\end{abstract}
\maketitle

%%%%%%%%%%%%%%%%%%%%%%%%%%%%%%%%%%%%%%%%%%%%%%%%%%%%%%%%%%%%%%%%%%%%%
%% Start the main part of the manuscript here.
%%%%%%%%%%%%%%%%%%%%%%%%%%%%%%%%%%%%%%%%%%%%%%%%%%%%%%%%%%%%%%%%%%%%%

One of the most common methods for preparing conventional optical vortex (COV) modes is imprinting the helical phase pattern onto the fundamental Gaussian mode through a spatial light modulator (SLM) \cite{a1} or a spiral phase plate (SPP) \cite{a2}. The COV beams have found important applications in a variety of fields such as improved image edge detection \cite{a7} and optical tweezers for manipulating particles \cite{a6} due to its unique phase structure. The most extensively researched topic regarding COV modes is high-dimensional communication \cite{a3,a4,a5} due to its potential for encoding in an infinite-dimensional Hilbert space. As for this topic, the implementation of high-dimensional entangled states \cite{a8}, frequency conversion \cite{a25} of COV beams \cite{a9,a24} and quantum memory for superposed COV modes \cite{a26,a10} have been realized recently.

However, the intrinsic dependence of the ring diameter of COV modes on the topological charge number $l$ limits its applications in scenarios where multiple modes with different $l$ are coupled into an optical system simultaneously. To overcome this obstacle, various concepts of structured light fields such as perfect Laguerre\mbox{-}Gaussian mode \cite{a13,a14} and flat-top beam have been proposed. For example, the frequency conversion of a 5-dimensional superposition state has been reported by using flat-top beams \cite{a15}. The most widely used kind of size-invariant light field is the perfect optical vortex (POV) beam proposed by Ostrovsky et al. \cite{a11}. A POV beam can be generated by Fourier transforming the corresponding Bessel-Guassian (BG) beam\cite{a12,a18}. It has been proved that POV beams offer advantages in establishing higher-dimensional quantum states over COV beams \cite{a16}, and POV states with different $l$ can also be distinguished and quantitatively identified in a projective measurement \cite{a17}. Although many pioneering works regarding the generation \cite{a18} or property analysis \cite{a19} of POV beams have been reported, and also its applications in optical manipulation \cite{a20,a21}, the high-dimensional frequency conversion of POV beams still remains to be a meaningful topic that needs to be studied. Here, we report a high-dimensional frequency conversion through the four-wave mixing (FWM) process in a hot atomic system. Our solutions can also be applied in a cold atomic system and thus it is useful for high-capacity and long-distance quantum communication.

\begin{figure*}[htp]
	\centering
	\includegraphics[width=430pt]{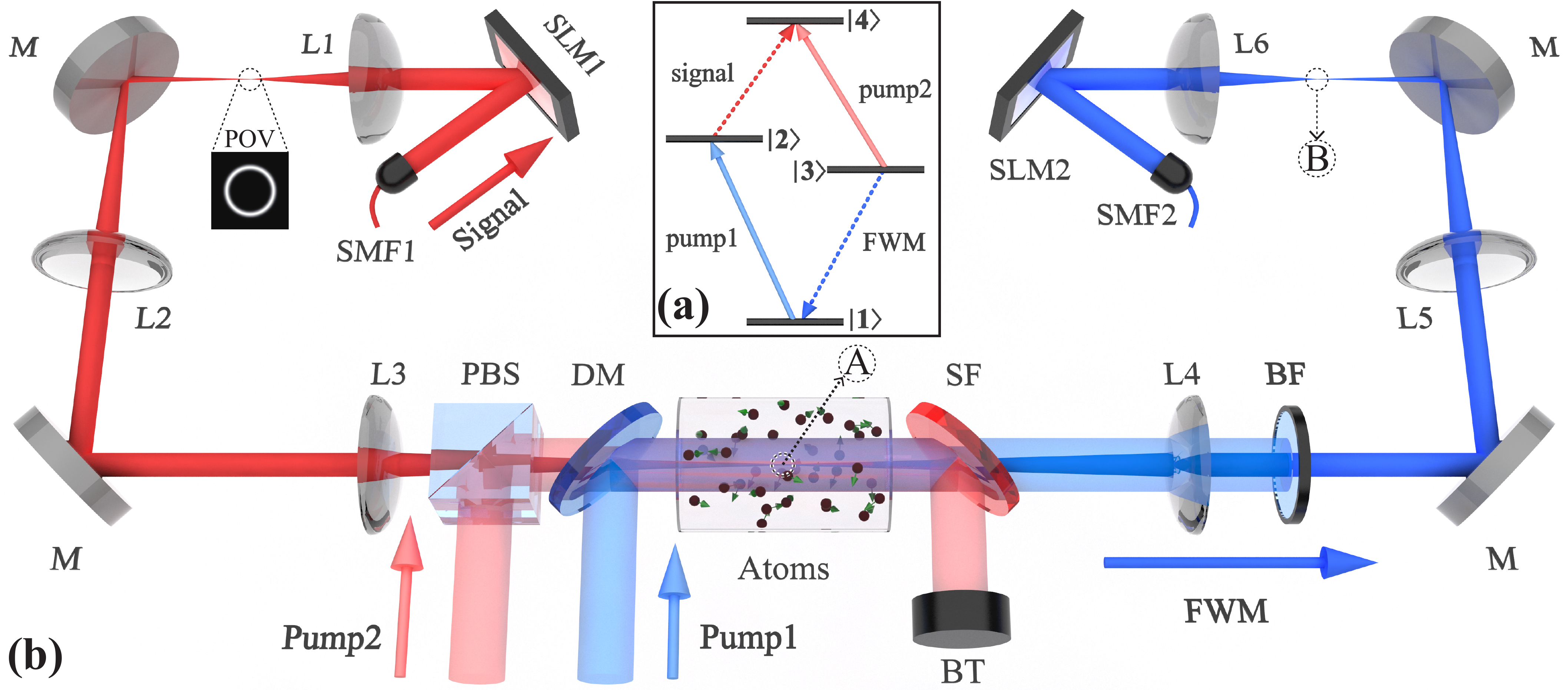}
	\caption{(a) Energy diagram of diamond configuration. (b) Schematic diagram of the experimental setup. SLM1, SLM2: spatial light modulator; PBS: polarizing beam splitter; The focal lengths of lenses L1, L2, L3, L4, L5 and L6 are 75, 150, 150, 150, 150, 75 mm, respectively; DM: long-pass dichroic mirror; SF: short-pass filter; BF: band-pass filter; BT: beam traps. SMF1, SMF2: single mode fiber.}
	\label{fig1}
\end{figure*}

As shown in Fig.\ref{fig1}(b), the POV beams in our experiment are generated by Fraunhofer diffracting a BG beam embedded in the corresponding helical phase. The latter is prepared by passing a fundamental Gaussian beam through a SLM with hologram $Arg\left[J_{l}(k_{r}r) e^{il\theta}\right]$ on it, here $Arg[\cdots]$ represents finding the argument, $J_{l}$ is $l$th order Bessel function of the first kind, $r$ and $\theta$ are radial and azimuthal coordinate respectively, $k_{r}=2.405/r_{0}$ is the radial wave vector with $r_{0}$ being the central core spot waist of the BG beam with $l=0$ \cite{a23}. The generated BG beam can be expressed as\cite{a22}:
\begin{equation}
E_{BG}(r,\theta)=J_{l}(k_{r}r)exp(\frac{-r^{2}}{\omega^{2}})exp(il\theta),
\label{eq:1}
\end{equation}
where $\omega$ is the waist of the original fundamental Gaussian beam. The lens L1 works as a Fourier transform system to obtain The POV beam, and it can be written as \cite{a12}
\begin{equation}
E_{POV}(r,\theta)=i^{l-1}\frac{\omega}{\omega_{0}}exp(il\theta)
exp(-\frac{r^{2}+r^{2}_{r}}{\omega^{2}_{0}})I_{l}(\frac{2r_{r}r}{\omega^{2}_{0}}),
\label{eq:2}
\end{equation}
where $\omega_{0}=2f/k\omega$ is the Gaussian beam waist at the rear focal plane of L1. $r_{r}=k_{r}f/k$ is the ring radius of the POV beam and $k=2\pi/\lambda$ is the wave vector. $I_{l}$ is an $l$th order modified Bessel function of the first kind.

The diamond-type energy configuration of $^{85}$Rb atom used in our experiment is shown in Fig.\ref{fig1}(a), which consists of one ground state $\left\lvert 1\right\rangle$ ($\left\lvert5S_{1/2},F=3\right\rangle$), one excited state $\left\lvert4\right\rangle$ ($\left\lvert 4D_{3/2},F^{\prime\prime}=3\right\rangle$) and two intermediate states $\left\lvert2\right\rangle$ ($\left\lvert5P_{3/2},F^{\prime}=3\right\rangle$) and $\left\lvert3\right\rangle$ ($\left\lvert 5P_{1/2},F^{\prime}=3\right\rangle$). The pump1 (780 nm), pump2 (1475 nm) and signal (1530 nm) lights couple the atomic transitions of $\left\lvert1\right\rangle \rightarrow \left\lvert2\right\rangle$, $\left\lvert3\right\rangle \rightarrow \left\lvert4\right\rangle$ and $\left\lvert 2\right\rangle \rightarrow \left\lvert4\right\rangle$ under resonance, respectively. According to the phase-matching condition of wave-vector conservation and energy conservation, a FWM light at 795 nm can be generated in the transition $\left\lvert1\right\rangle \rightarrow \left\lvert3\right\rangle$ through the FWM process.

The experimental setup is illustrated in Fig.\ref{fig1}(b). We prepare the POV beam (signal) by SLM1 and L1, and image it into the center of a 5-cm-long $^{85}$Rb vapor cell through a 4f system consisting of L2 and L3. The vertically polarized pump2 and vertically polarized pump1 lights are combined with the horizontally polarized signal light through a polarizing beam splitter (PBS) and a long-pass dichroic mirror (DM), respectively. The pump1 light and pump2 light propagate collinearly with the signal beam in the cell and have a waist of 1.67 mm (pump1) and 1.50 mm (pump2). The atomic cell is heated to 80 $^{\circ}$C to ensure a sufficiently high optical depth. In order to improve the signal-to-noise ratio, we use a short-pass filter and a band-pass filter to filter out the generated FWM light from the strong pump2 (100 mW) and pump1 (50 mW) lights respectively. Another 4f imaging system consisting of L4 and L5 images the frequency up-converted POV light (at point A) to point B for further detection with a charge coupled device (CCD). Finally, we perform the projective measurements with a lens L6, a SLM (SLM2) and a single mode fiber (SMF2) that is placed next to the second 4f imaging system. The FWM light collected via SMF2 is measured by a photomultiplier tube (PMT). The signal light (80 $\mu $W when it is continuous wave) is modulated to a square-shaped pulse with a temporal width of 1 $\mu$s.

\begin{figure}[htbp]
	\centering
	\includegraphics[width=8.5cm]{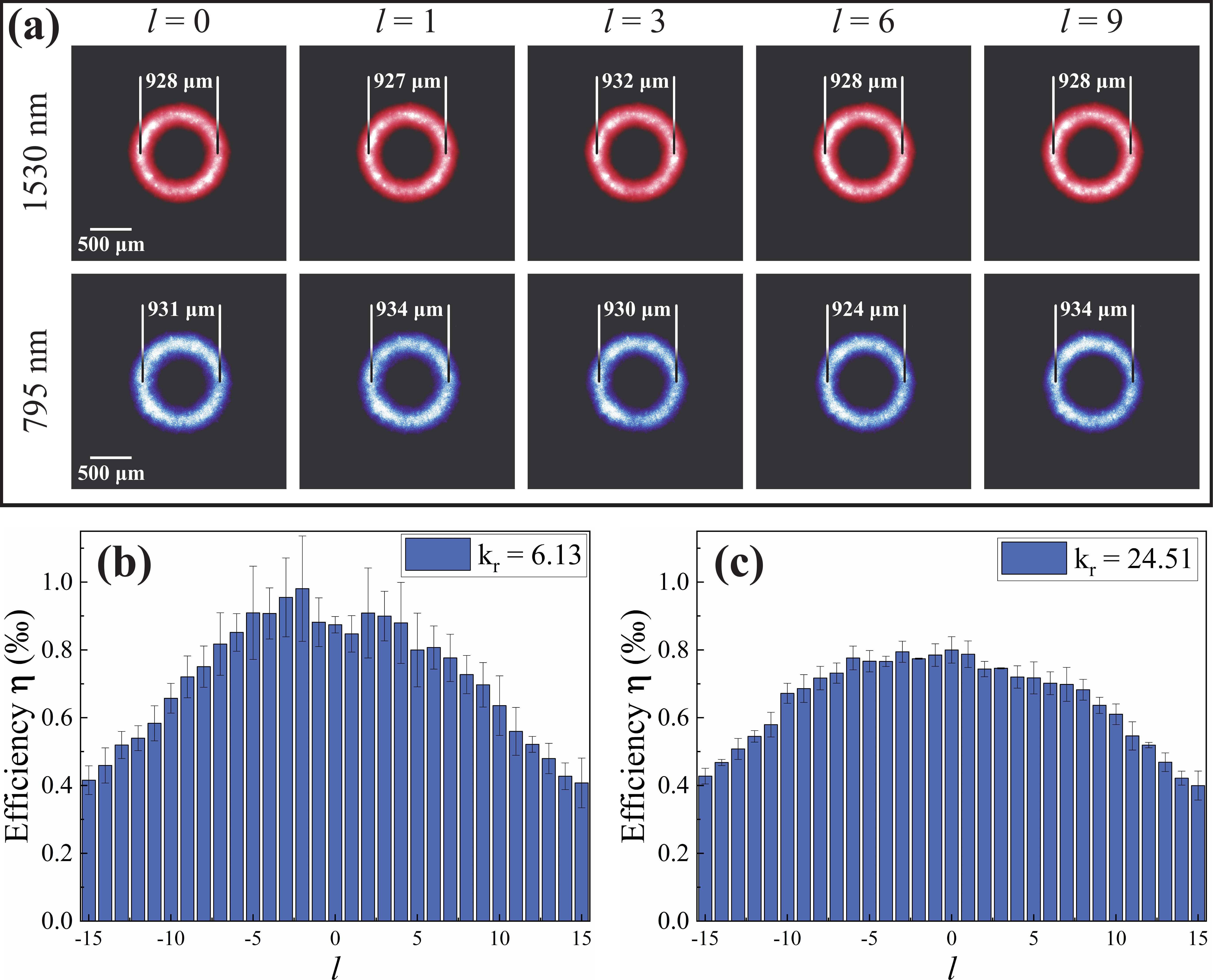}
	\caption{(a) The intensity profiles of input POV beam (red) and converted POV beam (blue). (b), (c) The distribution of conversion efficiency $\eta$ with different $l$ in the case of $k_{r}=6.13$ mm$^{-1}$ and $k_{r}=24.51$ mm$^{-1}$.}
	\label{fig2}
\end{figure}

We acquire the intensity profiles of the input (1530 nm) and converted (795 nm) POV beams through a CCD at the focal points A and B marked in Fig.\ref{fig1}(b), respectively. The results are shown in Fig.\ref{fig2}(a), both of the ring diameters of the two beams are calculated to be around 930 $\mu$m, which indicates a spatial-mode-conserved frequency conversion. What's more, this $l$-independent of ring diameter is consistent with the theory. The Fig.\ref{fig2} (b) and (c) exhibit the achieved conversion efficiency $\eta$ with different $l$ and $k_{r}$. Due to the existence of the last term in right hand side of the Eq.\ref{eq:2}, which originates from the Gaussian-shaped intensity distribution, the ring diameter will increase slowly with $l$ and this growth tendency becomes slower with the increase in $k_{r}$. The overlapped area between two pump beams and the POV beam changes with the increasing ring diameter, which leads to a reduced effective power of pump light that participates in the FWM process and thus a decreased $\eta$. We also find that the decreasing trend of $\eta$ becomes slower as $k_{r}$ increases by comparing Fig.\ref{fig2} (b) and (c). In this work, $\eta$ barely changes when $l$ is in the range of -6 to 6 and $k_{r}$ is 24.51 mm$^{-1}$, as shown in Fig.\ref{fig2}(c).

\begin{figure}[htbp]
	\centering
	\includegraphics[width=8.5cm]{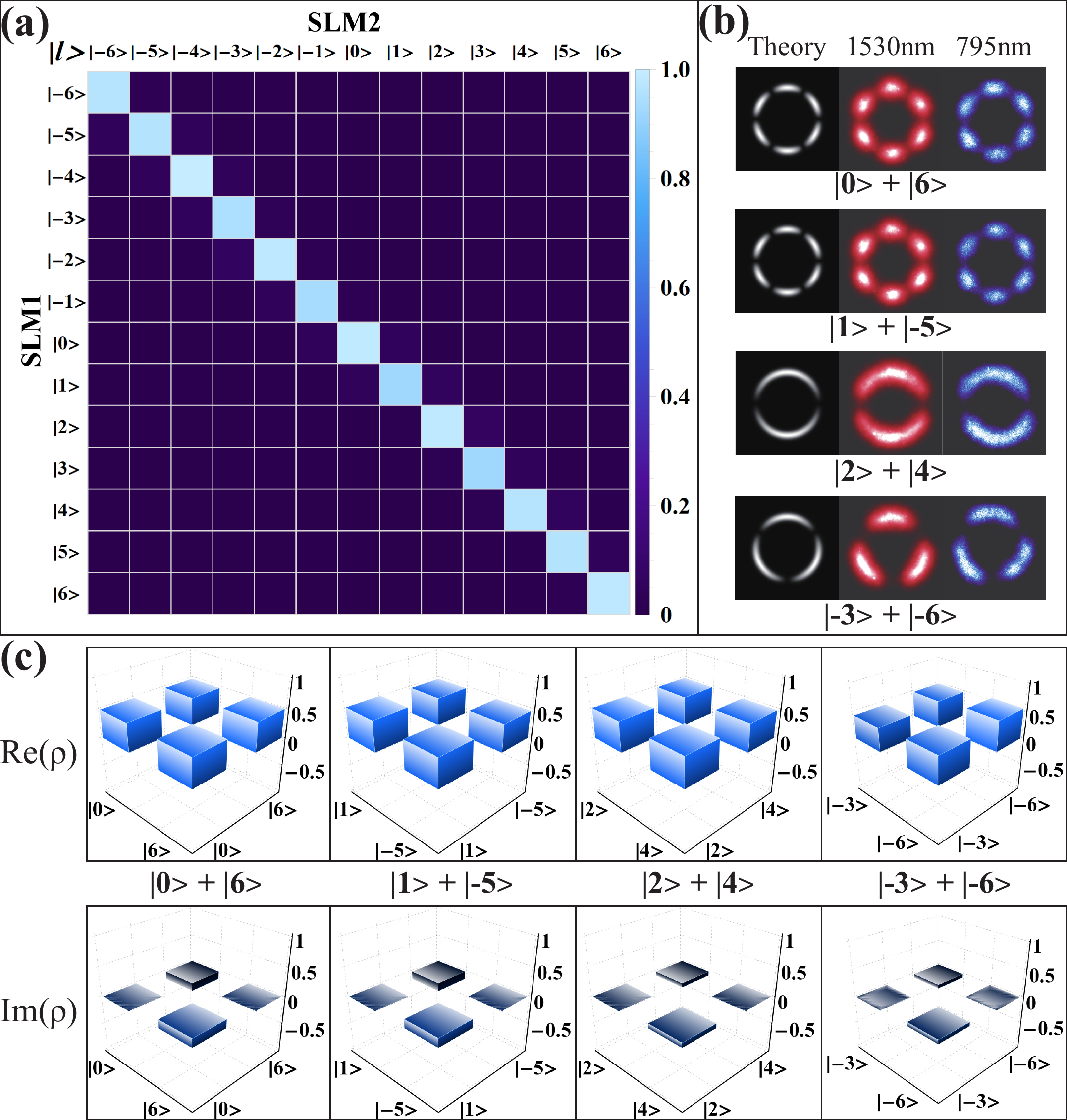}
	\caption{(a) Cross-talk matrix between input and converted beams formed by POV states in subspace $\left\{\left\lvert -6\right\rangle,...,\left\lvert 6\right\rangle\right\}$. (b) The theoretical, input and converted beam intensity profiles of four 2D states. (c) The real and imaginary parts of the reconstructed density matrix for the four 2D states.}
	\label{fig3}
\end{figure}

Fig.\ref{fig3}(a) depicts the normalized cross-talk matrix between input and converted POV beam, where the input and detected states $\left\lvert l\right\rangle$ are tailored from $\left\lvert -6\right\rangle$ to $\left\lvert 6\right\rangle$ by loading the corresponding phase hologram on SLM1 and SLM2 respectively. We define a SNR ($C=\sum_{a}M_{a,a}/\sum_{a,b}M_{a,b}$) of the cross-talk matrix to quantify the performance of our convertor. A high SNR of $C=90.97\pm0.23\%$ reveals that different POV states are well-distinguished from each other and have a low cross-talk noise. Because of the $l$-independent efficiency $\eta$ in the range of $l\in[-6,6]$, we are able to realize a high-dimensional frequency conversion for a state $\left\lvert\Psi\right\rangle$ consisting of arbitrary POV states $\left\lvert l\right\rangle$ within this subspace. Generally speaking, a N-dimensional (ND) superposition state can be written as 
\begin{equation}
\left\lvert\Psi\right\rangle=\frac{1}{\sqrt{N}}\sum_{i=1}^{N}\left\lvert l_{i}\right\rangle
\label{eq:3}
\end{equation}
with $l_{i}\in[-6,6]$.

We testify the frequency convertor with four 2D states that has diffierent values of $\Delta l(=\left\lvert l_{2}\right\lvert-\left\lvert l_{1}\right\lvert=6,4,3,2)$. These four states are listed as follows: $\left\lvert\Psi_{1}\right\rangle=(\left\lvert0\right\rangle+\left\lvert6\right\rangle)/\sqrt{2}$, $\left\lvert\Psi_{2}\right\rangle=(\left\lvert1\right\rangle+\left\lvert-5\right\rangle)/\sqrt{2}$, $\left\lvert\Psi_{3}\right\rangle=(\left\lvert2\right\rangle+\left\lvert4\right\rangle)/\sqrt{2}$ and $\left\lvert\Psi_{1}\right\rangle=(\left\lvert-3\right\rangle+\left\lvert-6\right\rangle)/\sqrt{2}$, Fig.\ref{fig3}(b) shows the theoretical profile, the registered profiles of input and converted beam of each state simultaneously. The generated beam profile is in good agreement with the theoretical simulation, and a high similarity between the intensity profiles of input and converted beams implies a faithful conversion process. We calculate the conversion fidelity ($F=Tr[\sqrt{\sqrt{\rho}\rho_{0}\sqrt{\rho}}]^{2}$) between the input and converted field by performing a projective measurement, here $\rho_{0}$ and $\rho$ are the theoretical and experimental density matrices respectively. The chosen bases for the measurement are $\left\lvert l_{1}\right\rangle$, $\left\lvert l_{2}\right\rangle$, $(\left\lvert l_{1}\right\rangle-i\left\lvert l_{2}\right\rangle)/\sqrt{2}$ and $(\left\lvert l_{1}\right\rangle+\left\lvert l_{2}\right\rangle)/\sqrt{2}$. The measured fidelities of $\left\lvert\Psi_{1}\right\rangle$, $\left\lvert\Psi_{2}\right\rangle$, $\left\lvert\Psi_{3}\right\rangle$ and $\left\lvert\Psi_{4}\right\rangle$ are $97.90\pm2.11\%$, $97.70\pm1.83\%$, $99.37\pm0.82\%$ and $99.06\pm0.55\%$ respectively, and the reconstructed density matrices are shown in Fig.\ref{fig3}(c). Our results indicate that the capability of our system to convert a 2D state $\left\lvert l\right\rangle$ in subspace $\left\{\left\lvert -6\right\rangle,...,\left\lvert 6\right\rangle\right\}$ faithfully.

\begin{figure}[htbp]
	\centering
	\includegraphics[width=8.5cm]{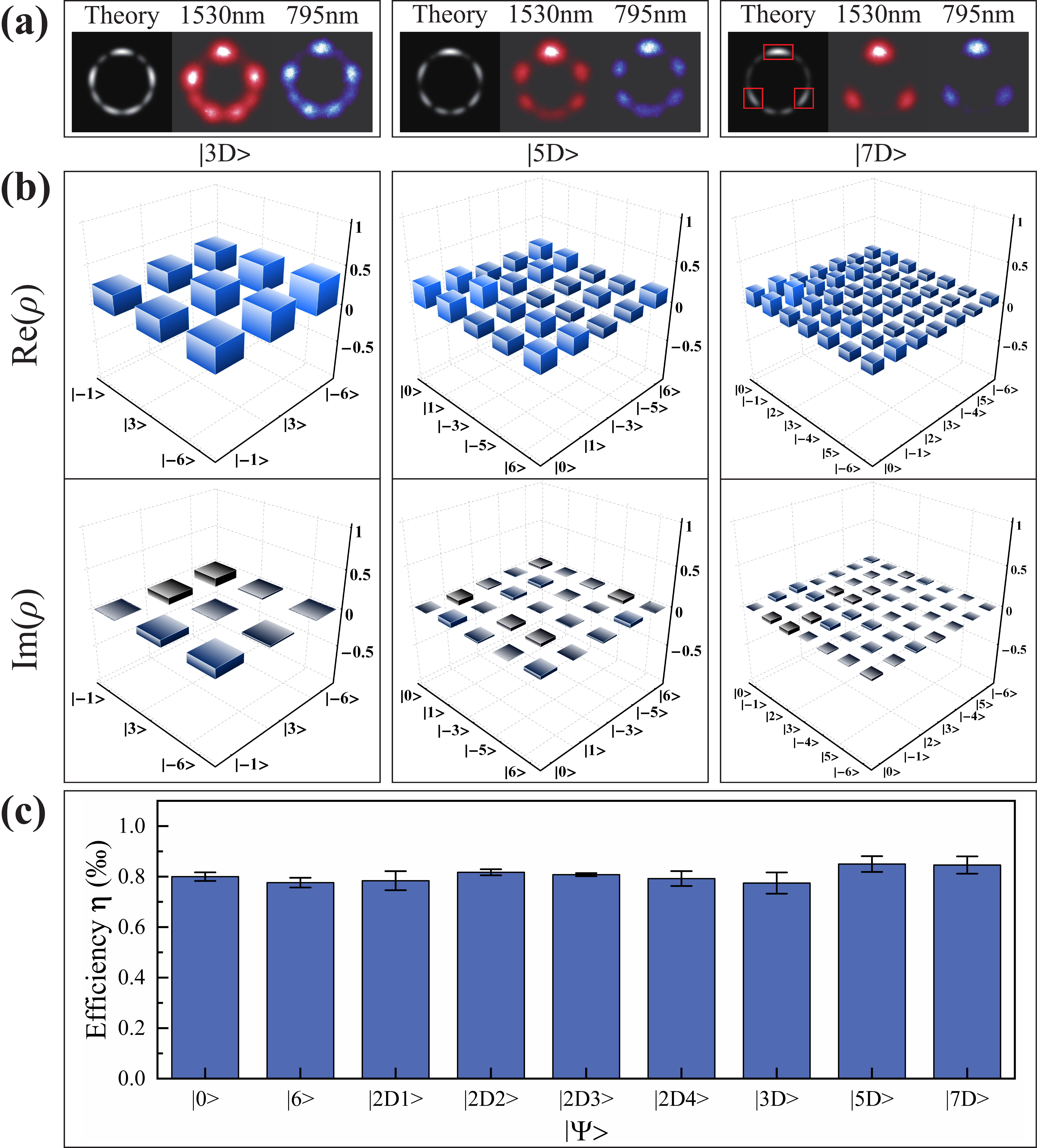}
	\caption{(a) The theoretical, input and converted beam intensity profiles of the 3D, 5D and 7D states. (b) The real and imaginary parts of the reconstructed density matrix for the 3D, 5D and 7D states. (c) The distribution of conversion efficiency $\eta$ with different states $\left\lvert\Psi\right\rangle$, where $\left\lvert2D1\right\rangle=(\left\lvert0\right\rangle+\left\lvert6\right\rangle)/\sqrt{2}$, $\left\lvert2D2\right\rangle=(\left\lvert1\right\rangle+\left\lvert-5\right\rangle)/\sqrt{2}$, $\left\lvert2D3\right\rangle=(\left\lvert2\right\rangle+\left\lvert4\right\rangle)/\sqrt{2}$, $\left\lvert2D4\right\rangle=(\left\lvert-3\right\rangle+\left\lvert-6\right\rangle)/\sqrt{2}$, $\left\lvert 3D\right\rangle=(\left\lvert-1\right\rangle+\left\lvert3\right\rangle+\left\lvert-6\right\rangle)/\sqrt{3}$, $\left\lvert 5D\right\rangle=(\left\lvert0\right\rangle+\left\lvert1\right\rangle+\left\lvert-3\right\rangle+\left\lvert-5\right\rangle+\left\lvert6\right\rangle)/\sqrt{5}$, and $\left\lvert 7D\right\rangle=(\left\lvert0\right\rangle+\left\lvert-1\right\rangle+\left\lvert2\right\rangle+\left\lvert3\right\rangle+\left\lvert-4\right\rangle+\left\lvert5\right\rangle+\left\lvert-6\right\rangle)/\sqrt{7}$.}
	\label{fig4}
\end{figure}

We then verify the validity of our system to work in 3D, 5D and 7D states by implementing the frequency conversion process of the following states: $\left\lvert 3D\right\rangle=(\left\lvert-1\right\rangle+\left\lvert3\right\rangle+\left\lvert-6\right\rangle)/\sqrt{3}$, $\left\lvert 5D\right\rangle=(\left\lvert0\right\rangle+\left\lvert1\right\rangle+\left\lvert-3\right\rangle+\left\lvert-5\right\rangle+\left\lvert6\right\rangle)/\sqrt{5}$, and $\left\lvert 7D\right\rangle=(\left\lvert0\right\rangle+\left\lvert-1\right\rangle+\left\lvert2\right\rangle+\left\lvert3\right\rangle+\left\lvert-4\right\rangle+\left\lvert5\right\rangle+\left\lvert-6\right\rangle)/\sqrt{7}$ respectively.  In Fig.\ref{fig4}(a), the complex intensity profile of the theoretically simulated beam leads to a weak intensity part of the beam that can't be detected by the CCD in our experiment, thus we observe a similarity between the theoretical and experimental beam profiles that tends to decrease as the dimensionality increases. However, the similarity between the input and converted beam profiles is still high. We chose the projected bases in the space of $\left\{\left\lvert l_{n}\right\rangle\right\}$, $\left\{\left\lvert l_{n}\right\rangle+\left\lvert l_{n+1}\right\rangle,...,\left\lvert l_{n}\right\rangle+\left\lvert l_{N}\right\rangle\right\}$, and $\left\{\left\lvert l_{n}\right\rangle+i\left\lvert l_{n+1}\right\rangle,...,\left\lvert l_{n}\right\rangle+i\left\lvert l_{N}\right\rangle\right\}$ with $n=1,...,N$, and make projective measurements to calculate the fidelity. The reconstructed density matrix is shown in Fig.\ref{fig4}(b) and the fidelities are $96.70\pm0.83\%$, $89.16\pm0.32\%$ and $88.68\pm1.23\%$ for 3D, 5D and 7D states. Due to the limited fixed pixel pitch (8 $\mu$m), the distortion of the hologram displayed by the SLM becomes more obvious as the phase pattern of the superposed POV state tends to be more complex with the increased number of dimensions. This result in differences in the detection efficiency of different bases during the projective measurements, and causes a decrease in measured fidelity. This may be overcome by using an SLM with higher pixel density or calibrating the experimental SLM2 before the measurement. By employing POV beams, we obtained a conversion fidelity over 88$\%$ even when the number of dimension reaches 7, in comparison, the achieved fidelity decreases to 50.97$\%$ for a 3D COV state ($(\left\lvert-1\right\rangle+\left\lvert3\right\rangle+\left\lvert-6\right\rangle)/\sqrt{3}$, note that the fidelity of the POV counterpart is 96.70$\%$). Our system provides advantages in extending the number of dimension of frequency conversion. Last but not the least, we have obtained a nearly unchanged conversion efficiency for superposed POV with different dimensions, as shown in Fig.\ref{fig4}(c). 

In conclusion, we report a high-dimensional frequency conversion in a hot atomic system through the FWM process. An $l$-independent frequency conversion process is achieved by using the POV beams. We find that the range of $l$ with constant conversion efficiency increases with the increasing $k_{r}$ used in our experiment, and verify that the capability of our system to convert a 2D superposition state in the subspace faithfully by performing frequency conversion on four 2D states with different value of $\Delta l$ and all of the measured fidelities exceed 97$\%$ after conversion. We finally perform the frequency conversion on 3D, 5D and 7D states, and find that the conversion fidelity reaches $88.68\pm1.23\%$ for the 7D state. Our scheme, where the conversion efficiency is $l$-independent, can be also compatible with a cold atomic system and may find applications in the field of high-dimensional and long-distance quantum communication.

\section*{Data availability statement}
The raw data supporting the conclusion of this article will be made available by the authors, without undue reservation.

\section*{Author contributions}
D-SD and B-SS coordianted the research project. W-HZ peformed the experimental fabrication, measurments and analyzed the data. All authors discussed the manuscript.

\section*{Conflict of interest}
The authors declare that the research was conducted in the absence of any commercial or financial relationships that could be construed as a potential conflict of interest.

\section*{Funding}
This work was supported by National Key R$\&$D Program of China (Grants No. 2017YFA0304800), Anhui Initiative in Quantum Information Technologies (Grant No. AHY020200), the National Natural Science Foundation of China (Grants No. U20A20218, No. 61722510, No. 11934013, No. 11604322, No. 12204461), and the Innovation Fund from CAS, the Youth Innovation Promotion Association of CAS under Grant No. 2018490, the Anhui Provincial Key Research and Development Project under Grant No. 2022b13020002, and the Anhui Provincial Candidates for academic and technical leaders Foundation under Grant No. 2019H208.

\bibliographystyle{jcp}
\bibliography{Hotatoms}

\end{document}